\documentclass[twocolumn]{aastex63}

\usepackage{amsmath}
\usepackage{ulem}
\usepackage{color}

\shorttitle{}
\shortauthors{Judge}
  
\begin{document}

\title{Magnetic connections across the  chromosphere-corona transition region}

\correspondingauthor{Philip Judge}
\email{judge@ucar.edu}
\author{Philip Judge}
\affiliation{High Altitude Observatory,
National Center for Atmospheric Research,
Boulder CO 80307-3000,
 USA } 

\newcommand\new[1]{#1}
\begin{abstract}

The plasma contributing to emission from the Sun 
between the cool chromosphere
($\le 10^4$K) and hot corona ($\ge 10^6$K) has been subjected to many different interpretations.  Here
we look at the magnetic structure 
of this transition region (TR) plasma, \new{based upon the implications of }
CLASP2 data of an active region
recently published by Ishikawa et al.,
and earlier IRIS and SDO data of quiet regions.  Ishikawa et al. found that large areas of
sunspot plages are magnetically unipolar as measured in the cores of  \ion{Mg}{2} resonance lines, formed in the lower transition region under low plasma-$\beta$ conditions. 
Here we show that IRIS images in the line cores 
have fibrils which well aligned with the overlying coronal loop segments 
seen in the 171 \AA{} channel of SDO. 
When the TR emission in active regions arise from plasma magnetically and thermally connected to
the corona, then  
the line cores
can provide 
the first credible magnetic boundary conditions for force-free 
calculations extended to the corona.
We also re-examine IRIS images of dynamic TR cool loops previously reported as a major contributor to 
transition region emission from the quiet Sun.  Dynamic cool loops contribute only a small fraction of the total TR emission from the quiet Sun. 
\end{abstract}

\keywords{Sun: atmosphere - Sun: chromosphere - Sun: transition region
  - Sun: corona - Sun: magnetic fields}

\section{Introduction}
\label{sec:introduction}

Ultraviolet emission from features 
formed between the $\approx 6000$ K  and $10^6$ K plasmas of 
the chromosphere and corona of the Sun have been studied quantitatively for 
over half a century 
\citep[e.g.][]{Detwiler+others1961,
Pottasch1964,
Burton+others1967}. 
%
The underlying chromosphere spans many pressure scale heights, it is 
a thermostat in that 
an increase in heating there leads 
largely to the increased energy of latent heat of ionization of hydrogen, 
and the accompanying freed electrons 
lead to rapid radiation losses through inelastic collisions with atoms and atomic ions. In contrast, the overlying  corona 
conducts heat more efficiently to lower temperature plasma than it can radiate, because abundant ions (particularly of H, He, C, N, O) are fully ionized, or belong to 
H- or He- like ions. Radiative  losses from these ions are modest, because almost as much energy is required to excite radiative transitions, requiring a change in 
principal quantum number, as it does to ionize them
(\citealp{Gabriel+Jordan1971}, reviewed by 
\citealp{Judge2019}).  Therefore,
any coronal heating which would otherwise raise coronal temperatures is instead rapidly ducted to plasma at lower temperatures. The radiation losses per particle peak near the middle of the transition region close to 10$^5$ K, where transitions with no changes of principal quantum
number, such as \ion{O}{6} and 
other members of, for example,  Li-, Be-, B-,   Na- like ions, 
efficiently emit UV radiation. 

Increased heating in the corona therefore leads to larger TR 
radiative losses predominantly at UV wavelengths, with minor 
increases in coronal temperature
\citep{Woolley+Allen1950}. This
physical scenario leads to  well-known ``scaling laws'' for coronal loops close to equilibrium conditions, which are non-linear in electron temperature 
\citep{McWhirter+Thoneman+Wilson1975, Rosner+others1978}.

The solar TR plasmas, when  in thermal contact with  both chromosphere and corona,  
are therefore sandwiched between two stable thermal reservoirs, but in itself 
tends to be unstable to modest perturbations. 
%
Consequently, little plasma can exist for long at intermediate temperatures
near $10^5$K
\citep{Mariska1992}.  
For example 
in 1D empirical models, 
TR plasma
spans of order 100 km in height, compared with  1500 km for the stratified 
chromosphere and many thousands of kilometers in the corona \citep[e.g., see the review by][]{Jordan1992}.
Further, it has been known for 7+ decades that static equilibria 
are impossible by consideration of radiative and classical conductive energy transport alone \citep{Giovanelli1949}.

TR radiation  generally emerges in the vacuum UV ($\lambda > 1100$ \AA) where normal incidence optics can be used, and in this region many 
space-based instruments have operated since the 1960s.
As a consequence of the above considerations, the TR has been a natural focus of observers to focus on the dynamics and overall behavior of TR plasma as it spectacularly reacts to modest changes in conditions above and below. 
To the observer then, the TR appears very dynamic, akin to the 
motions at the end of a whip, in response to drivers above and below.    The dynamic
spectra of TR lines
\citep[e.g.][]{Dere+Bartoe+Brueckner1989}
have received an unusual amount of
attention compared with
their relatively 
benign chromospheric and coronal counterparts  
\citep{Mariska1992}. 

\begin{table*}
\caption{IRIS 1400\AA{} slitjaw image sequences}
\label{tab:sji}
\begin{tabular}{llrrrrrrr}
\hline
\hline
Start time UT & N$_{exp}$ & X & Y & exp. &cadence & $\overline {DN/sec}$ & $\overline I$ & SSN\\
\hline
2013-10-02 06:47:23.5$^\ast$ &  100 &     1 &   948 &     2.0 &    11.8 &    19.5 & 2093& 107\\
2013-12-09 23:33:46.8$^\ast$ &  128 &   970 &    30 &     8.0 &    18.9 &    75.8 & 2038& 108\\
2013-09-13 06:36:51.6 &  300 &    -2 &     7 &     2.0 &     3.8 &    24 & 2560& 105\\
2013-12-01 00:42:38.7 &  550 &    -3 &     3 &     2.0 &     3.8 &    15.3 & 1643& 108\\
2017-04-03 07:40:01.0 &  640 &    -2 &     4 &    15.0 &    16.9 &    17.1 &  245& 25\\
2019-07-05 09:39:36.5 &  590 &     2 &    -6 &     8.0 &    11.2 &     4.2 &  112& 4\\
2019-07-08 09:49:21.4 &  600 &     2 &    -6 &     8.0 &    11.2 &     4.4 &  118 &4\\
2019-07-11 17:19:23.4 &  128 &     1 &    -2 &     8.0 &    37.4 &     3.2 &   85&4 \\
\hline
\end{tabular}\\
\vskip 6pt
$^\ast$These are two of the datasets examined by \cite{2014Sci...346E.315H}.
N$_{exp}$ is the number of slitjaw frames obtained, X and Y are coordinates of the
center of the frames in arcsec, exposure time (exp.) and cadence are in seconds, the mean intensity $\overline I$ is 
in erg~cm$^{-2}$~s$^{-1}$~sr$^{-1}$. 
SSN is the monthly sunspot number compiled by Space Weather Services. 
\end{table*}

\subsection{The purpose of the present work}

In the light of new results 
of magnetic field measurements above the active corona reported by 
\cite{Ishikawa+others2021},
we 
re-open an old debate concerning
the magnetic and thermal structures that lead to TR emission.   There are two camps of thought.
\begin{enumerate}
    \item The TR plasma's bright emission is caused 
    by magnetic field-aligned transport  processes, including but  not limited to  classical heat conduction down from the corona.  
    \item The TR plasma's bright emission comes from an ``unresolved fine structure'' (henceforth ``UFS'') in which the coronal energy flux does not contribute significantly to the emission, at least for plasmas below $\approx 2\times10^4$K.  Instead the energy is supplied by 
    upward directed mechanical energy, deposited locally and advected and radiated away. 
\end{enumerate}

The consequences of resolving this debate extend beyond studies of the TR.
There is intense interest in understanding coronal physics as it pertains to the 
irradiation of interplanetary space with variable UV and X-ray radiation, as well as 
to the ejection of plasma and 
magnetic fields.  Lines known to form in TR plasma formed 
under scenario 1. 
can be used to infer  magnetic energy, configurations, and evolution in the overlying corona. TR emission arising from scenario 2. can say nothing about
evolving magnetic properties of the corona.

\begin{figure*}[ht]
\includegraphics[width=0.8\linewidth]{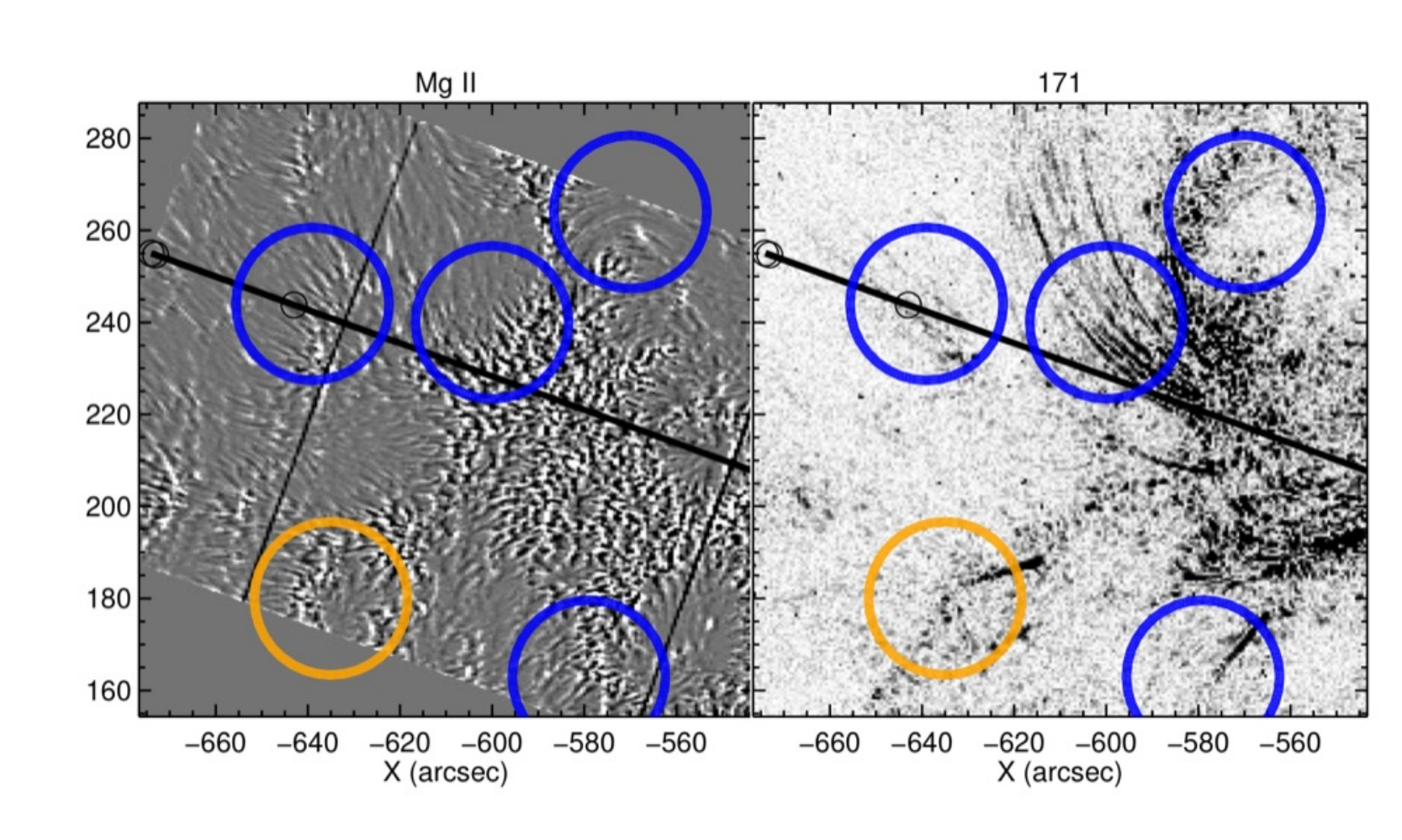}
\caption{Edge-enhanced images from near the core
of the Mg II k line (left panel, from an IRIS raster on
2019 April 11 from 15:57:54 to 16:25:26 UT)
and in the AIA 171 \AA{} channel (right panel, 16:53:33 UT) are shown.  The images are from
times close to the CLASP2 observations of \cite{Ishikawa+others2021}.
The central part of
these images is all of the same magnetic polarity as measured using CLASP2 data by \cite{Ishikawa+others2021} in the Mg II lines.  \new{The black line shows the
locus of the CLASP2 slit, the two small black circles correspond to the regions e and d of the chromospheric ``network'' 
identified by \cite{Ishikawa+others2021}. 
Large blue circles show areas where the 
chromospheric fibrils and coronal loops
appear well aligned, the orange circle 
shows a location where they are not.} 
}
\label{fig:slam}
\end{figure*}

\subsection{Field-aligned transport models}

In favor of models within camp 1., are two  facts related to the differential emission measure. This
well known quantity, derived by inversion of
observations, 
can be related to
the energy balance, including its transport, across the TR.
\cite{Jordan1980b}.
First the \textit{shape} of this function derived from observations of quiet Sun, coronal holes and active regions is remarkably constant \citep[e.g.][]{Noyes+Withbroe+Kuschner1970,Jordan1980b,Feldman+Dammasch+Landi2009}.  This
suggests a connection in the  transport of energy across the chromosphere-corona transition \citep{Jordan1992}. Secondly,
above $10^5$ K, the emission measure derived from observations 
varies as $T_e^{5/2}$ with $T_e$ the plasma electron temperature.   This result is
expected when energy is dominated by  field-aligned  heat conduction downwards
\citep[i.e. a constant conductive energy flux downwards, e.g.][]{Jordan1980b}. 
However, the emission measure structure below $10^5$ K is
leads to 
TR emission that is far brighter than classical models, based on heat conduction, could predict \citep{Kopp+Kuperus1968}.  
In principle this problem might be partly resolved 
in ``funnel-like'' expansion of the  magnetic field 
\cite{Gabriel1976}.  But in strong field concentrations where expansion has 
occurred below in the chromosphere, some have asked the question as to where the excess conducted energy 
goes \citep[e.g.][]{Kopp+Kuperus1968,Athay1990}.
The lack of static solutions identified first by \cite{Giovanelli1949}
has suggested to some that 
the excessively steep  
temperature gradients 
implied in classical 
models leads to dynamical instabilities (e.g., turbulence) which can then transport the 
conductive heat flux excess
across the atmosphere in a quasi-steady manner. The 
elementary plasma physics of the consequences of the high conductive flux has been carefully discussed by \citet{Ashbourn+Woods2001}.
A phenomenological formalism based 
on eddy diffusion was given by
\cite{Cally1990} in which the observed emission measures were used to find parameters of turbulent energy transport downwards to 10$^4$K. An alternate picture involving steady flows has been 
invoked by Fontenla and colleagues  who added neutral-ion diffusion processes (``ambipolar'' diffusion) to develop 
sophisticated models in 
a plane-parallel configuration \citep{Fontenla+Avrett+Loeser1990,Fontenla+Avrett+Loeser1993,Fontenla+Avrett+Loeser2002}. 
Some have suggested 
that spicules might arise 
from effects of 
this heat flux 
\citep{Kuperus+Athay1967,Kopp+Kuperus1968,
Athay2000}.

\citet{Athay1990} proposed
that field-aligned transport 
of conductive energy, when it
reaches cool plasma, may account for the emission measures below $10^5$ K. Athay invokes a highly 
corrugated (non-horizontal) 
thermal interface, generated perhaps  by the interface between the corona and embedded spicules. In this fashion,  
large cross-field temperature gradients may transport heat
from hot coronal ions into those in
the cooler plasma via ion-neutral collisions or anomalous cross-field 
transport processes 
\citep[see also][]{Ji+Song+Hu1996,Judge2008}.

\subsection{Models for the TR ``unresolved fine structures''}

In an intriguing series of 
papers, 
Feldman has argued that the 
bright TR emission observed 
is from structures 
which are energetically disconnected from the 
corona \citep{Feldman1983,Feldman1987,Feldman1998}.  Given that the 
thermal conductivities along the magnetic field are
many orders of magnitude 
greater than cross-field values at coronal temperatures, any bundles of
magnetic flux which never reach coronal temperatures 
will remain thermally protected from coronal plasma 
in which they may be embedded. Qualitative physical  interpretations by 
\cite{Dowdy+Rabin+Moore1986} 
 proposed that much of the quiet solar TR emission we see is not the result of downward energy transport.  Instead 
the structures are a population of ``cool loops'', with temperatures  
up to ca. $10^5$ K, near the peak of the radiative loss function, in which local 
mechanical heating from below leads to  the observed TR
radiative losses. \new{\cite{Antiochos+Noci1986} were able to relate the ideas to definite physical
models based upon physical force and energy balance. They 
argued that cool loops could, under reasonable conditions, 
explain long-recognized problems associated
only with classical heat transport along
hot loops. 
}
Various
authors studied the stability of such solutions  \citep[e.g.][]{Cally+Robb1991,Sasso+Andretta+Spadaro2015}, raising questions as to the
viability of the model.  However, this picture received support from the 
high resolution UV observations from the IRIS 
spacecraft, revealing dynamical 
closed loop-structures 
at the solar limb in spectral
features formed in the lower TR of the quiet Sun \citep[][henceforth ``H2014''  ]{2014Sci...346E.315H}.  The thermal evolution of the loops 
observed by \cite{2014Sci...346E.315H}
appears to vary
on timescales of a minute or so, comparable to sound crossing times.   Thus these
emitting structures are 
not static and instability may not
be an issue.  \new{In fact these structures are perhaps akin to the proposal of
\cite{Sasso+Andretta+Spadaro2015} 
in which populations of 
quasi-static models (those with subsonic speeds) might account for
a variety of important observations.} 

In the UFS picture, the question remains as yet unanswered: how can the 
emission measure distributions both below and above 10$^5$ K  be correlated, as observed, in this ``cool loop'' picture?  Presumably some kind of 
statistical averaging is invoked.  It is certainly true that the cool loops of H2014 are 
small, effectively below the resolutions of data used for most emission measure analyses, suggesting large numbers of small features will contribute over large areas. The question remains open.

Another question pertains to the 
nature of TR structures in 
\textit{active regions}. In plages surrounding sunspots, the TR emission is on average several times brighter than in quiet regions
\citep[e.g.][]{Brueckner+Bartoe1974}. 
Cool loops
seen during the SKYLAB era in active regions \citep[reviewed by][]{Jordan1976} 
might suggest
that these structures again 
dominate the TR emission.  These appeared most prominently under post-flare conditions.  Later results from the VAULT L$\alpha$ imager \citep{VAULT2007}
appeared to show cool loops
associated with plages surrounding active regions.
However, plages are known to
be largely unipolar \citep{Giovanelli1982}. Thus 
\citet{Judge+Centeno2008}
examined the underlying 
magnetic fields and the 
morphology of the L$\alpha$ images, concluding instead that these cannot be cool loops because plages  
underlying VAULT images are indeed
unipolar, as measured using 
sensitive 
magnetographs.  The authors suggested instead that the TR over plages forms through energy transport from both the chromosphere and corona. 

\section{Implications of data from the CLASP2 UV spectropolarimeter}

The first published results from the CLASP2 mission \citep{Ishikawa+others2021} focus on the Stokes $V$ profiles of
the resonance lines of
\ion{Mg}{2} as observed primarily in plages.  Their results are important for our study of the TR,
to include both active and quiet  regions.   While the bulk of the radiation in the opacity-broadened \ion{Mg}{2} $h$ and $k$  lines 
is emitted from the chromosphere 
\citep{Ayres1979}, the very cores 
form in plasma close to $2\times10^4$ K
\new{in 1D models of \cite{Fontenla+Avrett+Loeser1993} 
(Judge et al. in preparation).  These line cores can therefore probe magnetic conditions in the lower
TR plasma.  The calculations of line core
formation heights in the 3D dynamical snapshots of \cite{2013ApJ...772...90L}
are unfortunately not plotted as a function of electron temperature $T_e$.  But when the $h$ and $k$ lines become optically thin because of 
increased ionization to Mg$^{2+}$ by electron collisions and not simply the 
drop in opacity when Mg$^+$ is the dominant ion, then the $\tau=1$ surface must occur close
to $2\times10^4$ K.  The contours of 
the $\tau=1$ surfaces of \ion{Mg}{2} $k$ in Figure 1 of \cite{2013ApJ...772...90L} lie close to the $2\times10^4$ K temperature 
cutoff shown, seeming to confirm this picture.
}

\begin{figure*}[ht]
\includegraphics[width=\linewidth]{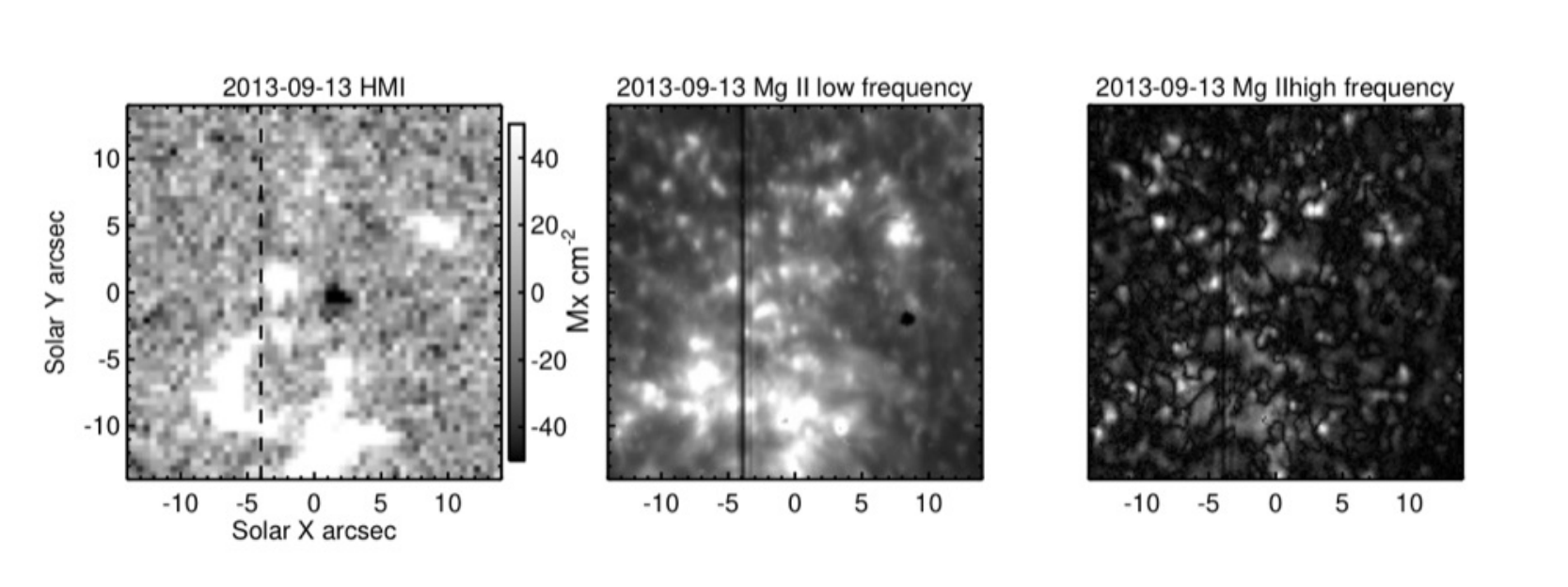}
\caption{The left panel shows 
an HMI magnetogram obtained several
minutes before the 
frequency-filtered  1400\AA{} slitjaw images in the right panels were obtained at disk center on 2013-09-13 near to sunspot maximum.  The middle panel shows the intensity 
in all features where the 
dominant power is at frequencies below 8 mHz (a period of 120 seconds).  The right power shows the absolute value of power at high frequencies ($>8$ mHz).  
The morphology is slightly different in
the HMI image because of the difference in timing of the order of tens of minutes between the HMI and IRIS timeseries of observations. The area shown is about the same as one supergranule.  This is an area considered as ``quiet'' even though 
it was obtained when the Sun was relatively active.
}
\label{fig:max}
\end{figure*}

\cite{Judge+Centeno2008} studied the morphology and the underlying photospheric magnetic environment of VAULT L$\alpha$ data.  
They argued that the cool loop solutions cannot be important 
as active region TR 
contributors for several reasons.  Firstly, the photospheric magnetic fields underlying the 
bright and curious L$\alpha$ structures in the images appear unipolar.  
The 
unipolar fields 
measured using the Kitt Peak Vacuum Telescope   (KPVT), had an average flux per resolution element of $\approx 1.4
\times 10^{18}$ Mx.  In contrast,  the smallest
detectable flux concentrations of opposite sign to the 
major polarity 
were 1.3$\times10^{16}$ Mx (a $2\sigma$ statistical level). 
Although it is possible to hide the magnetic footpoints of loops if they are 
below this level, the Sun would have to arrange to do so such that \textit{essentially no} flux of opposite polarity 
could be detected over an entire plage.  One might expect, based upon the disorder associated with 
magneto-convection, that patches of
opposite polarity, if present with sufficient flux, should be visible. 

\cite{Judge+Centeno2008} also argued that 
the morphology and 
spatial alignment of 
the L$\alpha$ ``fibrils'' arranged themselves more like little comets, with tails somehow aligned 
over large areas.  This situation is anything other that that envisaged for the quiet Sun \citep{Dowdy+Rabin+Moore1986}.

The CLASP2 data sample 
magnetic fields directly at the base of the solar transition region. Such  plasma is many  
orders of magnitude lower in pressure than,
and $\approx$2 Mm above,  the photosphere where previous magnetic measurements have been made. 
The Stokes $V$ profiles detected using CLASP2 in the observed plages and active network are 
\textit{all of the same sign}.  In any model 
of the Zeeman effect (\citealp{Ishikawa+others2021} interpreted the data quantitatively using the weak field approximation), this can only mean unipolar \new{line-of-sight (LOS)} fields.   The CLASP2 results are consistent 
with the physical picture of magnetic field decreasing in strength from their origins beneath the solar surface.

Following 
\cite{Judge+Centeno2008},
in Figure~\ref{fig:slam} we attempt to reveal possible alignments between the comet-like Mg II fibril structures at the edges of the plage
and in the network to the west, with overlying coronal structure.   A 1:1 correspondence between the two is not seen, but this is
as expected given the disparate lengths and 
heights of the two kinds of structures. The Mg II ``fibrils''
are seen only  at the edges of 
the plages. This is perhaps due to
the dimmer  background intensities 
over the quieter regions, and/or the denser populations and hence confusion expected over the
central area of plages.
Nevertheless, given the different projections of the chromospheric and coronal structures, the overall alignment is remarkable in the underlying organized directions along which magnetic structures
can be traced.   

We conclude that the evidence for the connection of bright plages and associated bright network emission in TR lines 
to the corona is 
definitive.  

\section{Analysis of IRIS 
data of the quiet Sun transition region}

Here we examine more IRIS data
from the 1400 \AA{} channel of
the Slit Jaw Imager (SJI).  We will conclude that indeed 
dynamic cool loops found by 
H2014 
contribute to the quiet Sun's
emission in the resonance lines
of \ion{Si}{4}, formed 
in the middle TR near  $10^{4.8}$ K.  However, we will also find that the 
statistics of dynamical cool loops identified by H2014 
indicate that these structures contribute far less that 50\%{} to the emission from the quiet Sun.  This stands in contrast to the 
claim by H2014:
\begin{quote}
     IRIS observations prove that a large fraction of the solar transition region emission is due to low lying, relatively cool, loops having no thermal contact with the corona.
\end{quote}

Firstly we examine the 
actual data analyzed by H2014, listed along with 
other data  listed in Table~\ref{tab:sji}.
The data of H2014 were data taken at the solar limb, from October and December 2013, a period near the maximum of  solar cycle 24.   From histograms of properties of these dynamic loops (their Figure 2), 
we adopt the following numbers for analysis. The median values of
apparent length, height above limb and intensity (in DN sec$^{-1}$) are
4 Mm, 2 Mm and $\approx 40$ DN sec$^{-1}$ respectively.   A distribution of lifetimes is not shown, but we assume that the time series shown in Figure S1 are typical, and estimate median lifetimes 
of order 60 seconds. This number is mentioned in their
text as 
\begin{quote}
    ...loops appear to be lit up in segments, with each segment only being visible for roughly a minute. 
\end{quote}

\begin{figure*}[ht]
\includegraphics[width=\linewidth]{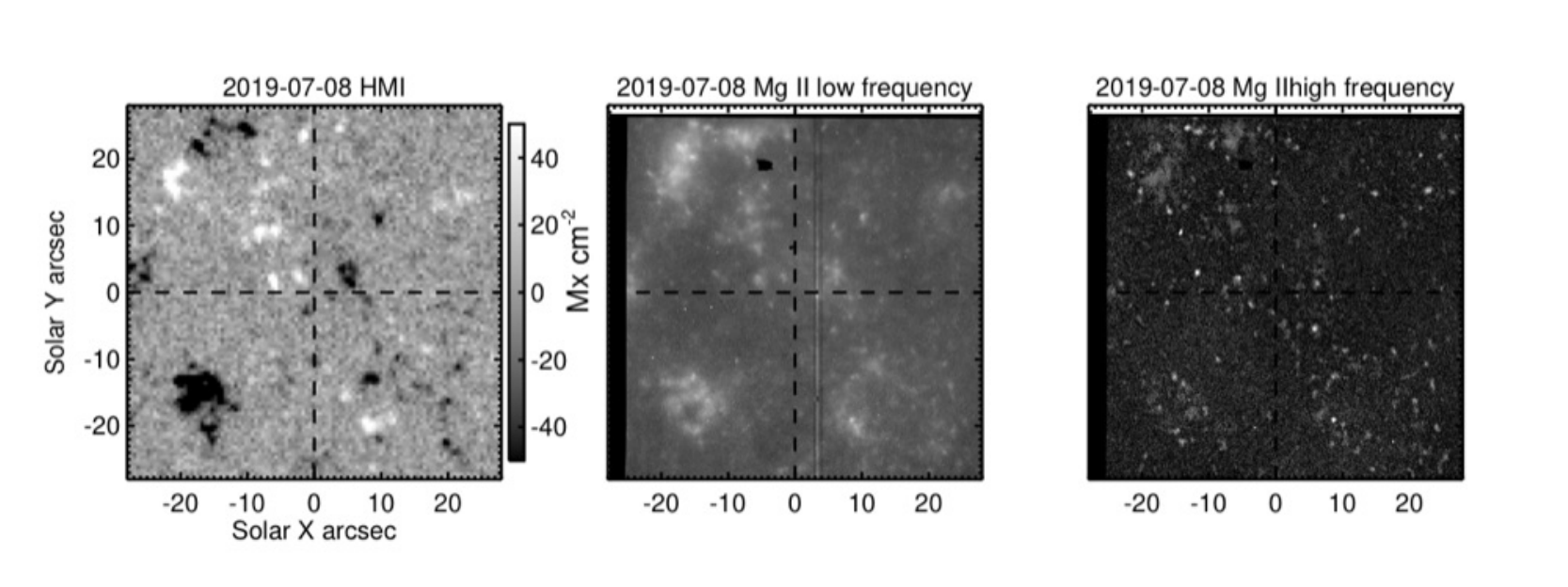}
\caption{  A figure identical 
to Figure~\ref{fig:max} except
that the data were acquired close to solar minimum conditions, it can be considered as ``very quiet'' Sun.   ``Cool loop'' (dipolar) contributions are present near the top left of each panel, other areas of emission are more unipolar. Color scales are the same as for 
Figure~\ref{fig:max}.  The lines help to reveal the precision of the alignment of the images.
}
\label{fig:min}
\end{figure*}

We will require two more statistical quantities- the number of cool loops visible at any time per unit arc length along the limb, and their apparent area in the plane of the sky.   These quantities are not specified in the original paper but inspection of the limb and inside-limb 
images (their Fig. 1) indicates to us that there are at most 3 such loops visible over the quiet Sun 
limb every arc-minute of arc length (i.e., one such loop every length $s\approx14$ Mm). While this number is subject to
large uncertainties given the 
published analysis, movie S1 of H2014 seems to confirm
this rough number. 
Certainly 
in the data shown, the number cannot be twice this value
for the regions shown.  
The loops shown in their Fig.~1 
indicate a median loop 
area $A$ of $\approx 4 \cdot 0.5=2$ Mm$^2$. 

Let us transform the limb loops to disk center in a statistical sense.  The 
geometry at the limb means that features of median height $h$ will be visible if it lies within a line-of-sight length \begin{equation}
\label{eq:geom}
\ell \approx \sqrt{2 R_\odot h} \approx 50 \ \mathrm{Mm}
\end{equation}
The equivalent area observed at disk center 
in the quiet Sun containing 
just 1 dynamic loop is then 
$\ell \cdot s \approx 700$ Mm$^2$.  The area of a single supergranule is roughly 700 
Mm$^2$. We should expect to observe one of these bright loops per supergranule 
at any given time. 

The total radiative flux 
from these features, assuming they are optically thin, is independent of viewing angle. 
Using the reported median intensity of UFS at the limb 
of $I=40$ DN~sec$^{-1}$ (H2014), we  
would compute a mean disk-center intensity
$\overline I$
distributed over area $\ell \cdot s$ of 
\begin{equation}
\overline I \approx AI/(\ell \cdot s)     \approx 0.003 I
\end{equation}
H2014 give a physical conversion 
of 215 erg cm$^{-2}$~sr$^{-1}$~s$^{-1}$ for each DN per second, so that $I \approx 8600$ erg~cm$^{-2}$~sr$^{-1}$~s$^{-1}$.  Taken together, we find that the summed intensity of all the limb cool loops when viewed at disk center
in the 1400\AA{} channel of the SJI is then
\begin{equation}
\label{eq:icen}
    \overline I \approx 26
\mathrm{ \  erg~cm^{-2}~sr^{-1}~s^{-1} }
\end{equation}
which includes both lines of \ion{Si}{4}.  Perhaps some of
the limb emission is hidden behind the extra opacity 
near 1400\AA{} along such long lines of sight. 
In the Discussion section below, we relate the limb
lines of sight to disk center
and address attenuation 
by bound-free opacity of neutral silicon. 

What might these structures then contribute to the 
quiet Sun intensities seen at disk center?  To avoid 
inter-calibration issues, 
we examined disk center 
data from IRIS acquired 
with the 1400\AA{} SJI channel listed in Table~\ref{tab:sji}.  These sample various levels of 
sunspot activity from 
nearly maximum sunspot number of over 100 down to 4 per month.   Clearly the mean disk center intensities vary greatly depending on the 
magnetic flux at disk center, for which the monthly SSN is a crude measure.  Even though Sun center is often 
used to observe typical ``quiet Sun'' conditions, in the IRIS SJI images there is no
typical value.
Nevertheless, every case we have examined has significantly higher intensities than estimated from equations (\ref{eq:geom}) --  (\ref{eq:icen}), even those at sunspot minimum which 
represent really quiet Sun.

Are these SJI intensities comparable to previous measurements of
``average quiet Sun'' intensities of TR lines at disk center?  The ``averages'' 
are rather poorly defined because the rms spatial variations in \ion{Si}{4} lines 
are on the order of the mean
values \citep{Athay+Dere1991}.  Nevertheless, the SUMER atlas  of the quiet Sun has $\overline I \approx 260$ erg~cm$^{-2}$sr$^{-1}$s$^{-1}$
\citep{Curdt+others2001}. Brekke's atlas of HRTS data 
has $\overline I 
\approx 170$ \citep{Brekke1993}.  All these
estimates are at least an order of magnitude 
larger than the summed contributions of the limb
dynamic loop intensities, using the above parameters.

To check  this analysis, 
we  filtered the 
disk center IRIS data in time using a filter 
centered at 120 seconds, twice the lifetime of the dynamics loops. 
The time series SJI frames analyzed here and by H2014 are between 3.8 and 18.9 seconds. 
Low-pass data were generated using a Gaussian filter to remove dynamic 
variations on time scales of
1 minute.    
The resulting 
low-pass time series was then 
subtracted from the initial data to produce a high-pass 
time series and the 
low- and high- bandpass datasets compared.  
Typical results for data obtained in 2013 
are shown in figure
\ref{fig:max}.   The data obtained 
near sunspot maximum, show features at low frequencies with intensities up to $2\times10^4$ erg cm$^{-2}$sr$^{-1}$ s$^{-1}$.  The high frequency component  
has features only as bright
as 800 erg cm$^{-2}$sr$^{-1}$s$^{-1}$.
Figure~\ref{fig:min} shows similar data obtained close to 
solar sunspot minimum (from Jul 8 2019 with a monthly sunspot number of 4).  This figure, covering abour 4 supergranular areas, has a good example
of what to expect from a cool loop (discussed below), near (X,Y=-17,+20).  

In both cases the features with significant power at periods below 120 seconds (right hand panels) account for about 25\% of the total
power (including zero frequency).

We conclude that using  
representative properties of
the dynamic loops identified by H2014, 
\textit{the dynamic cool loops
of H2014 are a small contributor to the total emission of the quiet Sun TR. }

\begin{figure*}[ht]
\includegraphics[width=0.9\linewidth]{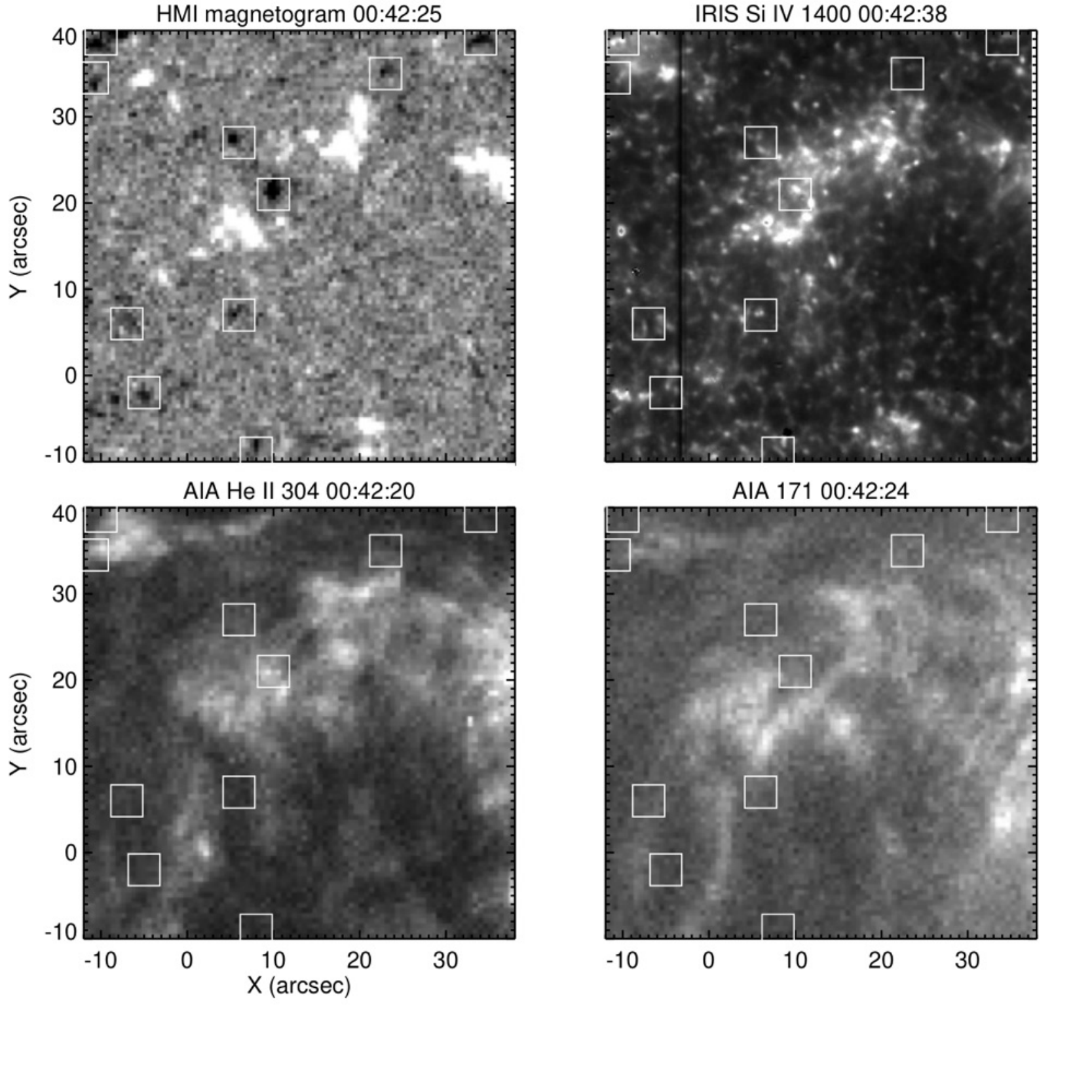}
\caption{Various data obtained on
December 1 2013 close in time to the first IRIS 
image of a sequence obtained at disk center.  The area shown is roughly that of three supergranules.
Synoptic views reveal this to be 
a quiet region, although with a 
monthly SSN of 108, the 1400 SJI data are quite bright 
compared with solar minimum conditions 
(Table~\ref{tab:sji}).  The images are co-aligned to within a second of arc. Some of the patches of 
negative polarity are marked with 
boxes for reference between images. The HMI magnetogram is saturated at $\pm 40$ Mx~cm$^{-2}$, the strongest flux concentrations range between -150 and +100 Mx~cm$^{-2}$.
}
\label{fig:dec}
\end{figure*}

\section{Discussion}

\subsection{The active solar  transition region}

The conclusions of the analyses of 
\cite{Judge+Centeno2008} and \cite{Ishikawa+others2021} appear 
unassailable.   The active region plages  consist of bright emission from with unipolar magnetic structures threading from chromosphere (or lower, perhaps)  into the overlying corona, as inferred from  Figure~\ref{fig:slam}.

Readers may be surprised in that the TR associated with active regions is itself
active (dynamic), with much evidence for
magnetic reconnection \citep[e.g.][]{Dere+Bartoe+Brueckner1989}.  However, it
should be remembered that reconnection will occur in unipolar magnetic fields which define 
the overall structure.  Within these unipolar fields, tangential discontinuities
\citep{Parker1988,Parker1994} are expected which can readily lead to reconnection not of the strong unipolar but the transverse components.  This conclusion will modify some conclusions in the extensive literature \citep[e.g.][and later work]{Mariska1992}.

The consequences of the work of \cite{Ishikawa+others2021} are particularly interesting for applications to understand the origin of flares, CMEs and other phenomena associated with the slow storage and sudden release of magnetic energy 
in the corona
\citep{Gold+Hoyle1960}. Measurements of 
magnetic fields in the lines of \ion{Mg}{2} 
can be used to probe the magnetic structure and evolution as active regions evolve and 
release magnetic energy, plasma, radiation and perhaps magnetic helicity into
interplanetary space.

\subsection{The ``quiet'' Sun}

The main results of our study of quiet regions at disk center 
are based upon  
Figures~\ref{fig:max} (a close-up of data from Sep 13 2013 with a sunspot number of 105) 
and \ref{fig:dec} (a larger view including TR and coronal AIA data from SDO).   For comparison, 
and because the Sun center 
intensities measured by IRIS are
a strong function of monthly sunspot number, Figure \ref{fig:min} shows similar data covering about 4 supergranule areas when the 
monthly sunspot number was 4.  
In all cases but one,
the brightest
1400 \AA{} emission lies directly above the HMI magnetograms 
of one polarity.  The exception is the bipolar region of Figure~\ref{fig:min} close to
(X,Y=-17,+20) where emission at both
high and low frequencies are seen spanning the opposite polarity regions. In the more active disk center data (Figure~\ref{fig:dec}), the lower two panels reveal coronal
structures: the He II channel formed near $10^5$ K shows structure clearly associated with the Fe IX/X 171 \AA{}
coronal brightness.  This again suggests an energetic connection of a unipolar region to the overlying
corona.  
The spatial relationships are far from 1:1 because 
the coronal emission is spread along
magnetic fields by efficient 
electron heat conduction.  

But to what extend might there be 
direct evidence that the corona and the IRIS 1400 \AA{} emission are  energetically connected?  The very different angular
resolutions (0.32" for IRIS vs. 1" for SDO) preclude identifying 
the small bright IRIS structures with
the overlying atmosphere. However, the 
white boxes in Figure~\ref{fig:dec} highlight some of the stronger regions of
opposite polarity. Two such boxes
(X=-6,Y=7 and X=11,Y=22) also include nearby
positive polarity and so low-lying small loops might connect these regions.  Indeed there is a small 
amount of Si IV 1400 \AA{} emission present but this does not span
from positive to negative polarity.
In these data there is therefore 
little indication of cool loop emission. This is as we expected on the 
basis of statistics of limb data 
obtained during the same season
derived above (perhaps just one dynamic cool loop per supergranular area). 

Our results for quiet regions 
appear to be self-consistent, with the 
possible exception of the 
N. polar sequence of frames
acquired on 2013-10-02, with 
mean intensities equal to those of the scan on 2013-12-09 at lower latitudes.   Perhaps these bright 
loops
are related to the roots of 
extended polar plumes which are magnetically
multi-polar, relatively dense and themselves bright.  

Here we find that the magnetic and IRIS data indicate that quiet Sun TR emission is dominated by plasma connected to the corona which, under typical coronal conditions, must lie near 2000 km above the 
photosphere.  Spicular emission in TR lines will appear dominant at the limb because they are long and 
via equation~(\ref{eq:geom})
they are
many times brighter than at disk center. 
This then raises the question, 
why then  is the proposed bright TR plasma connected
to the corona not seen everywhere at the solar limb, for example in the data from Figure~1 of H2014?

We recall that
absorbers at 1400 \AA{} are formed near heights of 0.8 Mm where $\tau=1$ 
vertically at disk center (opacity from neutral Si, \citealp{Vernazza+Avrett+Loeser1981}).  At the limb, we can 
find a lower limit to the projected height at which the $\tau=1$ surface will arise assuming the atmosphere to be
exponentially stratified. The density  scale height $h_c$ in the mid
chromosphere of
  \cite{Vernazza+Avrett+Loeser1981} is about 150 km.  
  Then, the limb path length tangential to the limb is 
  $\ell \approx \sqrt{2R_\odot h_c}$ is about 15 Mm, 100 times $h_c$
  (equation \ref{eq:geom}).
  The $\tau=1$ surface will be found at a projected high $z$ above the disk center height of 800 km, when \begin{eqnarray}
    \ell \cdot  \exp(-z/h_c)&=& 
    1,\ \ \mathrm{ so\  that} \\ \nonumber
    z &=& 1400 \ \mathrm{km},
\end{eqnarray}
However, in these 1D models this height is 
limited not only by the height dependence of the density, 
but also by temperature 
since the coronal temperature rises to 1 million K near heights of 2000 km at disk center. Thus $z$ is perhaps  at most 
1200 km in these models, and so 
any limb emission at heights $\le 800+1200 = 2000$ km will be attenuated.  Given the 
presence of dynamic extensions higher than the hydrostatic scale heights 
(clearly shown in Figure~1 of \citealp{Judge2015} based upon data from the Bifrost code of \citealp{Gudiksen+others2011}), this is perhaps
a lower limit. So, this is a minimal estimate of the
attenuation of 1400 \AA{} radiation at the limb.  Brighter TR emission lies below the 2000 km 
height (see the 
scaling laws for coronal 
pressure of \citealp{McWhirter+Thoneman+Wilson1975,Rosner+others1978}
and the sequence of 
models A-F of \citealp{Vernazza+Avrett+Loeser1981} for example). Not only this, but the presence of spicules and other 
cool material  and 
the dynamic nature of the chromosphere in 
simulations means that 
the bulk of the bright TR emission at the footpoints of the corona will be mostly absent in the images analyzed by H2014.

There is qualitative agreement of the visibility of loops observed with IRIS and those in the numerical work of H2014.  However, the initial mixed polarity state used in these calculations maximizes the number of low-lying loops.
Therefore these modeling efforts cannot be used to refute our 
conclusions. 

\new{Finally, CLASP2 data analyzed by \cite{Ishikawa+others2021} include regions of network (points d and e marked in their Figure~1 and Figure 3, and in this paper 
as small black circles in Figure~\ref{fig:slam}). The Zeeman signals seen in the \ion{Mg}{2} $h$ and $k$ line cores at these two locations are compatible with zero LOS magnetic
field. The authors showed that these 
points are of opposite polarity in the photosphere with LOS average flux densities of $\approx 220$ Mx~cm$^{-2}$.   Outside of
the cores, the \ion{Mg}{2} lines yielded
160 and 80 Mx~cm$^{-2}$ LOS flux densities
in the lower and middle chromosphere
respectively.  \cite{Ishikawa+others2021} 
speculate that the non-detection may arise
because associated loops may lie beneath
the formation height of the line cores, 
because the magnetic flux densities 
lie below the Zeeman detection limit
of $\approx 10$ Mx~cm$^{-2}$, or because
the fields are perpendicular to the LOS.  
Here we can rule out the last possibility
because local verticals at these 
points are inclined at $\approx 43^\circ$ 
to the LOS, and any locally horizontal connecting loop structure between them would have to 
have as strong Zeeman component along the LOS as a vertical component. Until higher quality data and/or 
inversion scheme can be acquired, the precise meaning of these network observations for 
the conclusions of the present paper will remain unclear.  
}

In conclusion, the bulk of transition region emission from quiet and active regions on the Sun does not arise from UFS, but from a nagnetically- guided thermal interface between
the corona and the chromosphere. This interface is probably highly 
corrugated, geometrically (see Figure~\ref{fig:slam}, also 
\citealp[e.g.][]{Athay1990,Ji+Song+Hu1996,Judge2008}).

\acknowledgments
The author is grateful to 
Rebecca Centeno Elliott for reading the paper and for encouragement, and to an anonymous referee for very helpful comments.
The National Center for Atmospheric Research is sponsored by the National Science Foundation.  


\bibliography{biblio}

\begin{thebibliography}{}
\expandafter\ifx\csname natexlab\endcsname\relax\def\natexlab#1{#1}\fi
\providecommand{\url}[1]{\href{#1}{#1}}
\providecommand{\dodoi}[1]{doi:~\href{http://doi.org/#1}{\nolinkurl{#1}}}
\providecommand{\doeprint}[1]{\href{http://ascl.net/#1}{\nolinkurl{http://ascl.net/#1}}}
\providecommand{\doarXiv}[1]{\href{https://arxiv.org/abs/#1}{\nolinkurl{https://arxiv.org/abs/#1}}}

\bibitem[{{Antiochos} \& {Noci}(1986)}]{Antiochos+Noci1986}
{Antiochos}, S.~K., \& {Noci}, G. 1986, {ApJ}, 301, 440, \dodoi{10.1086/163912}

\bibitem[{{Ashbourn} \& {Woods}(2001)}]{Ashbourn+Woods2001}
{Ashbourn}, J.~M.~A., \& {Woods}, L.~C. 2001, Proceedings of the Royal Society
  of London Series A, 457, 1873, \dodoi{10.1098/rspa.2001.0791}

\bibitem[{Athay(1990)}]{Athay1990}
Athay, R.~G. 1990, ApJ, 362, 364

\bibitem[{{Athay}(2000)}]{Athay2000}
{Athay}, R.~G. 2000, Sol. Phys., 197, 31

\bibitem[{{Athay} \& {Dere}(1991)}]{Athay+Dere1991}
{Athay}, R.~G., \& {Dere}, K.~P. 1991, {ApJ}, 379, 776, \dodoi{10.1086/170553}

\bibitem[{Ayres(1979)}]{Ayres1979}
Ayres, T.~R. 1979, ApJ, 228, 509

\bibitem[{{Brekke}(1993)}]{Brekke1993}
{Brekke}, P. 1993, {ApJ}S, 87, 443, \dodoi{10.1086/191810}

\bibitem[{{Brueckner} \& {Bartoe}(1974)}]{Brueckner+Bartoe1974}
{Brueckner}, G.~E., \& {Bartoe}, J.-D.~F. 1974, Sol. Phys., 38, 133

\bibitem[{{Burton} {et~al.}(1967){Burton}, {Ridgeley}, \&
  {Wilson}}]{Burton+others1967}
{Burton}, W.~M., {Ridgeley}, A., \& {Wilson}, R. 1967, MNRAS, 135, 207,
  \dodoi{10.1093/mnras/135.2.207}

\bibitem[{Cally(1990)}]{Cally1990}
Cally, P.~S. 1990, ApJ, 355, 693

\bibitem[{Cally \& Robb(1991)}]{Cally+Robb1991}
Cally, P.~S., \& Robb, T.~D. 1991, ApJ, 372, 329

\bibitem[{{Curdt} {et~al.}(2001){Curdt}, {Brekke}, {Feldman}, {Wilhelm},
  {Dwivedi}, {Sch{\"u}hle}, \& {Lemaire}}]{Curdt+others2001}
{Curdt}, W., {Brekke}, P., {Feldman}, U., {et~al.} 2001, A\&A, 375, 591

\bibitem[{Dere {et~al.}(1989)Dere, Bartoe, \&
  Brueckner}]{Dere+Bartoe+Brueckner1989}
Dere, K.~P., Bartoe, J.-D.~F., \& Brueckner, G.~E. 1989, {ApJ}, 345, L95

\bibitem[{{Detwiler} {et~al.}(1961){Detwiler}, {Garrett}, {Purcell}, \&
  {Tousey}}]{Detwiler+others1961}
{Detwiler}, C.~R., {Garrett}, D.~L., {Purcell}, J.~P., \& {Tousey}, R. 1961,
  Annales de Geophysique, 17, 263

\bibitem[{{Dowdy} {et~al.}(1986){Dowdy}, {Rabin}, \&
  {Moore}}]{Dowdy+Rabin+Moore1986}
{Dowdy}, J.~F., J., {Rabin}, D., \& {Moore}, R.~L. 1986, Sol. Phys., 105, 35

\bibitem[{Feldman(1983)}]{Feldman1983}
Feldman, U. 1983, ApJ, 275, 367

\bibitem[{Feldman(1987)}]{Feldman1987}
---. 1987, ApJ, 320, 426

\bibitem[{{Feldman}(1998)}]{Feldman1998}
{Feldman}, U. 1998, ApJ, 507, 974

\bibitem[{{Feldman} {et~al.}(2009){Feldman}, {Dammasch}, \&
  {Landi}}]{Feldman+Dammasch+Landi2009}
{Feldman}, U., {Dammasch}, I.~E., \& {Landi}, E. 2009, {ApJ}, 693, 1474,
  \dodoi{10.1088/0004-637X/693/2/1474}

\bibitem[{Fontenla {et~al.}(1990)Fontenla, Avrett, \&
  Loeser}]{Fontenla+Avrett+Loeser1990}
Fontenla, J.~M., Avrett, E.~H., \& Loeser, R. 1990, ApJ, 355, 700

\bibitem[{Fontenla {et~al.}(1993)Fontenla, Avrett, \&
  Loeser}]{Fontenla+Avrett+Loeser1993}
---. 1993, ApJ, 406, 319

\bibitem[{{Fontenla} {et~al.}(2002){Fontenla}, {Avrett}, \&
  {Loeser}}]{Fontenla+Avrett+Loeser2002}
{Fontenla}, J.~M., {Avrett}, E.~H., \& {Loeser}, R. 2002, ApJ, 572, 636

\bibitem[{Gabriel(1976)}]{Gabriel1976}
Gabriel, A. 1976, Phil Trans. Royal Soc. Lond., 281, 339

\bibitem[{Gabriel \& Jordan(1971)}]{Gabriel+Jordan1971}
Gabriel, A.~H., \& Jordan, C. 1971, Case Studies in Atomic Collision Physics,
  ed. E.~McDaniel \& M.~C. McDowell (North-Holland), 210--291

\bibitem[{{Giovanelli}(1949)}]{Giovanelli1949}
{Giovanelli}, R.~G. 1949, MNRAS, 109, 372, \dodoi{10.1093/mnras/109.3.372}

\bibitem[{{Giovanelli}(1982)}]{Giovanelli1982}
---. 1982, Sol. Phys., 77, 27, \dodoi{10.1007/BF00156093}

\bibitem[{{Gold} \& {Hoyle}(1960)}]{Gold+Hoyle1960}
{Gold}, T., \& {Hoyle}, F. 1960, MNRAS, 120, 89

\bibitem[{{Gudiksen} {et~al.}(2011){Gudiksen}, {Carlsson}, {Hansteen}, {Hayek},
  {Leenaarts}, \& {Mart{\'{\i}}nez-Sykora}}]{Gudiksen+others2011}
{Gudiksen}, B.~V., {Carlsson}, M., {Hansteen}, V.~H., {et~al.} 2011, A\&A, 531,
  A154

\bibitem[{{Hansteen} {et~al.}(2014){Hansteen}, {De Pontieu}, {Carlsson},
  {Lemen}, {Title}, {Boerner}, {Hurlburt}, {Tarbell}, {Wuelser}, {Pereira}, {De
  Luca}, {Golub}, {McKillop}, {Reeves}, {Saar}, {Testa}, {Tian}, {Kankelborg},
  {Jaeggli}, {Kleint}, \& {Mart{\'\i}nez-Sykora}}]{2014Sci...346E.315H}
{Hansteen}, V., {De Pontieu}, B., {Carlsson}, M., {et~al.} 2014, Science, 346,
  1255757, \dodoi{10.1126/science.1255757}

\bibitem[{Ishikawa {et~al.}(2021)Ishikawa, Bueno, del Pino~Alem{\'a}n, Okamoto,
  McKenzie, Auch{\`e}re, Kano, Song, Yoshida, Rachmeler, Kobayashi, Hara, Kubo,
  Narukage, Sakao, Shimizu, Suematsu, Bethge, De~Pontieu, Dalda, Vigil,
  Winebarger, Ballester, Belluzzi, {\v S}t{\v e}p{\'a}n, Ramos, Carlsson, \&
  Leenaarts}]{Ishikawa+others2021}
Ishikawa, R., Bueno, J.~T., del Pino~Alem{\'a}n, T., {et~al.} 2021, Science
  Advances, 7, \dodoi{10.1126/sciadv.abe8406}

\bibitem[{{Ji} {et~al.}(1996){Ji}, {Song}, \& {Hu}}]{Ji+Song+Hu1996}
{Ji}, H.~S., {Song}, M.~T., \& {Hu}, F.~M. 1996, {ApJ}, 464, 1012,
  \dodoi{10.1086/177388}

\bibitem[{Jordan(1976)}]{Jordan1976}
Jordan, C. 1976, Phil. Trans. R. Soc. London, A281 , 391

\bibitem[{Jordan(1980)}]{Jordan1980b}
---. 1980, A\&A, 86, 355

\bibitem[{{Jordan}(1992)}]{Jordan1992}
{Jordan}, C. 1992, Memorie della Societa Astronomica Italiana, 63, 605

\bibitem[{{Judge} \& {Centeno}(2008)}]{Judge+Centeno2008}
{Judge}, P., \& {Centeno}, R. 2008, {ApJ}, 687, 1388, \dodoi{10.1086/590104}

\bibitem[{Judge(2008)}]{Judge2008}
Judge, P.~G. 2008, ApJL, 683, L87

\bibitem[{{Judge}(2015)}]{Judge2015}
{Judge}, P.~G. 2015, ApJ, 808, 116

\bibitem[{{Judge}(2019)}]{Judge2019}
---. 2019, {Chapter 5 - Spectroscopy and Atomic Physics}, ed. O.~{Engvold},
  J.-C. {Vial}, \& A.~{Skumanich}, 127--155,
  \dodoi{10.1016/B978-0-12-814334-6.00005-4}

\bibitem[{{Kopp} \& {Kuperus}(1968)}]{Kopp+Kuperus1968}
{Kopp}, R.~A., \& {Kuperus}, M. 1968, Sol. Phys., 4, 212

\bibitem[{{Kuperus} \& {Athay}(1967)}]{Kuperus+Athay1967}
{Kuperus}, M., \& {Athay}, R.~G. 1967, Sol. Phys., 1, 361,
  \dodoi{10.1007/BF00151361}

\bibitem[{{Leenaarts} {et~al.}(2013){Leenaarts}, {Pereira}, {Carlsson},
  {Uitenbroek}, \& {De Pontieu}}]{2013ApJ...772...90L}
{Leenaarts}, J., {Pereira}, T.~M.~D., {Carlsson}, M., {Uitenbroek}, H., \& {De
  Pontieu}, B. 2013, ApJ, 772, 90, \dodoi{10.1088/0004-637X/772/2/90}

\bibitem[{Mariska(1992)}]{Mariska1992}
Mariska, J.~T. 1992, The Solar Transition Region (Cambridge UK: Cambridge
  Univ.\ Press)

\bibitem[{McWhirter {et~al.}(1975)McWhirter, Thoneman, \&
  Wilson}]{McWhirter+Thoneman+Wilson1975}
McWhirter, R. W.~P., Thoneman, P.~C., \& Wilson, R. 1975, A\&A, 40, 63

\bibitem[{{Noyes} {et~al.}(1970){Noyes}, {Withbroe}, \&
  {Kirshner}}]{Noyes+Withbroe+Kuschner1970}
{Noyes}, R.~W., {Withbroe}, G.~L., \& {Kirshner}, R.~P. 1970, Sol. Phys., 11,
  388, \dodoi{10.1007/BF00153074}

\bibitem[{Parker(1988)}]{Parker1988}
Parker, E.~N. 1988, ApJ, 330, 474

\bibitem[{Parker(1994)}]{Parker1994}
---. 1994, Spontaneous Current Sheets in Magnetic Fields with Application to
  Stellar X-Rays, International Series on Astronomy and Astrophyics (Oxford:
  Oxford University Press)

\bibitem[{{Patsourakos} {et~al.}(2007){Patsourakos}, {Gouttebroze}, \&
  {Vourlidas}}]{VAULT2007}
{Patsourakos}, S., {Gouttebroze}, P., \& {Vourlidas}, A. 2007, {ApJ}, 664,
  1214, \dodoi{10.1086/518645}

\bibitem[{Pottasch(1964)}]{Pottasch1964}
Pottasch, S.~R. 1964, Sp. Sci. Rev., 3, 816

\bibitem[{Rosner {et~al.}(1978)Rosner, Tucker, \& Vaiana}]{Rosner+others1978}
Rosner, R., Tucker, W.~H., \& Vaiana, G.~S. 1978, {ApJ}, 220, 643

\bibitem[{{Sasso} {et~al.}(2015){Sasso}, {Andretta}, \&
  {Spadaro}}]{Sasso+Andretta+Spadaro2015}
{Sasso}, C., {Andretta}, V., \& {Spadaro}, D. 2015, A\&A, 583, A54,
  \dodoi{10.1051/0004-6361/201526598}

\bibitem[{Vernazza {et~al.}(1981)Vernazza, Avrett, \&
  Loeser}]{Vernazza+Avrett+Loeser1981}
Vernazza, J., Avrett, E., \& Loeser, R. 1981, ApJS, 45, 635

\bibitem[{{Woolley} \& {Allen}(1950)}]{Woolley+Allen1950}
{Woolley}, R.~V.~D.~R., \& {Allen}, C.~W. 1950, MNRAS, 110, 358,
  \dodoi{10.1093/mnras/110.4.358}

\end{thebibliography}
\bibliographystyle{aasjournal}

\end{document}